\documentclass[runningheads]{llncs}
\usepackage{verbatim}
\usepackage{amsmath}
\usepackage{amssymb}
\usepackage{enumitem}
\usepackage{marvosym} 

\usepackage{graphicx}
\usepackage{subcaption}

\usepackage[T1]{fontenc}
\usepackage{booktabs}
\usepackage{graphicx,verbatim}
\begin{document}
\title{HaDM-ST: Histology-Assisted Differential Modeling for Spatial Transcriptomics Generation }
%
\begin{comment}  %% Removed for anonymized MICCAI 2025 submission
\author{First Author\inst{1}\orcidID{0000-1111-2222-3333} \and
Second Author\inst{2,3}\orcidID{1111-2222-3333-4444} \and
Third Author\inst{3}\orcidID{2222--3333-4444-5555}}
%
\authorrunning{F. Author et al.}
\institute{Princeton University, Princeton NJ 08544, USA \and
Springer Heidelberg, Tiergartenstr. 17, 69121 Heidelberg, Germany
\email{lncs@springer.com}\\
\url{http://www.springer.com/gp/computer-science/lncs} \and
ABC Institute, Rupert-Karls-University Heidelberg, Heidelberg, Germany\\
\email{\{abc,lncs\}@uni-heidelberg.de}}

\end{comment}

\author{
Xuepeng Liu$^{2,*}$ \and
Zheng Jiang$^{2,*}$ \and
Pinan Zhu$^{2}$ \and
Hanyu Liu$^{2}$ \and
Chao Li$^{1}$\textsuperscript{\Letter}
}

\institute{
$^1$ University of Cambridge, UK \\
Correspondence author (\Letter): \texttt{cl647@cam.ac.uk}\\
$^2$ Northeastern University, Shenyang, China \\
\texttt{\{20237370, 20246365, 20237359\}@stu.neu.edu.cn}, 2485644@dundee.ac.uk 
}

\authorrunning{X. Liu et al.}
\titlerunning{HaDM-ST: Histology-Assisted Differential Modeling for ST Generation}

\maketitle

\renewcommand{\thefootnote}{}
\footnotetext{* These authors contributed equally to this work.}
\renewcommand{\thefootnote}{\arabic{footnote}}  % 若后续仍需正常脚注

\begin{abstract}
Spatial transcriptomics (ST) reveals spatial heterogeneity of gene expression, yet its resolution is limited by current platforms. Recent methods enhance resolution via H\&E-stained histology, but three major challenges persist: (1) isolating expression-relevant features from visually complex H\&E images; (2) achieving spatially precise multimodal alignment in diffusion-based frameworks; and (3) modeling gene-specific variation across expression channels. We propose \textbf{HaDM-ST} (Histology-assisted Differential Modeling for ST Generation), a high-resolution (HR) ST generation framework conditioned on H\&E images and low-resolution (LR) ST. HaDM-ST includes: (i) a semantic distillation network to extract predictive cues from H\&E; (ii) a spatial alignment module enforcing pixel-wise correspondence with low-res ST; and (iii) a channel-aware adversarial learner for fine-grained gene-level modeling. Experiments on 200 genes across diverse tissues and species show HaDM-ST consistently outperforms prior methods, enhancing spatial fidelity and gene-level coherence in HR ST predictions.


\end{abstract}
\section{Introduction}

Spatial transcriptomics (ST) has revolutionized our understanding of tissue biology by providing spatially resolved gene expression. However, the spatial resolution of most mainstream ST platforms remains inherently limited \cite{7}, as they typically measure gene expression at coarse, spot-level granularity, which hinders fine-scale spatial analysis. Although recent high-resolution (HR) ST technologies such as Xenium \cite{1} and Visium \cite{6} emerge, they are costly and often suffer from reduced capture efficiency \cite{8}, limiting their real-world applications.

To overcome these limitations, recent efforts have explored the potential of leveraging histology context, particularly hematoxylin-and-eosin (H\&E) stained tissue sections, to infer HR ST data and improve its spatial resolution. Among the generative modeling approaches, conditional diffusion models have emerged as a powerful solution in medical image synthesis tasks \cite{4,10}. These models simulate a denoising Markov chain conditioned on auxiliary inputs, enabling the generation of realistic HR images from low-resolution (LR) or multimodal inputs. Their inherent stochasticity allows for uncertainty-aware prediction, while their conditioning mechanisms provide flexibility to integrate diverse sources of biological information \cite{2,3,5}.

In this study, we explore a cross-modal generation paradigm in which HR ST maps are synthesized by integrating H\&E histology morphology with corresponding LR ST measurements. Unlike conventional super-resolution methods that merely upscale existing ST data, our approach learns a modality translation process guided by histology context and augmented by transcriptomic priors. Specifically, the LR ST provides coarse-grained gene expression levels across spatial regions, along with gene–gene co-expression relationships, serving as a biological prior that informs both expression intensity and inter-gene structural dependencies during generation.

To effectively leverage the histological morphology for ST generation, three core challenges remain:
\noindent\textbf{Complex histology semantics:} H\&E images contain rich and heterogeneous visual features, making it difficult to isolate expression-relevant morphological cues that correlate with gene activity;
\noindent\textbf{Multi-conditional misalignment:} Traditional diffusion pipelines struggle to align heterogeneous modalities, such as histology textures and transcriptomic signal, at pixel-level precision, especially when conditioned on coarse-resolution ST inputs;
\noindent\textbf{Lack of gene-specific modeling:} ST data consists of multiple gene expression channels, each reflecting unique biological patterns. Existing methods lack mechanisms to explicitly model gene-wise variations across these channels.

To address these challenges, we propose \textbf{HaDM-ST} (Histology-assisted Differential Modeling for ST Generation), a diffusion-based image translation framework that generates HR ST maps from H\&E images, guided by LR ST inputs during training. Our method introduces three key innovations.

\begin{itemize}
    \item \textbf{H\&E-Driven Semantic Distillation (HSD):} A transformer-based semantic encoder that filters out irrelevant histology noise and distills expression-relevant features from H\&E morphology.
    
    \item \textbf{Cross-Modal Spatial Alignment (CMSA):} A pixel-level alignment module based on contrastive learning, which uses LR ST data to guide the alignment between histology and transcriptomic features.
    
    \item \textbf{Gene-wise Differential Adversarial Learning (GDAL):} A graph-based gene modeling module that incorporates a channel-aware discriminator to capture inter-gene relationships and refine gene-specific expression in the predicted ST maps.
\end{itemize}

Extensive experiments across 200 genes from public ST datasets covering multiple tissues and species demonstrate that HaDM-ST consistently outperforms existing approaches, achieving superior spatial fidelity and gene-level accuracy in the generated HR ST outputs.

\section{Related Work and Problem Statement}
ST is rapidly evolving from spot–based sequencing toward subcellular and even single-cell imaging \cite{11}. High sequencing costs and resolution bottlenecks, however, still hinder its widespread clinical adoption. A growing body of research leverages readily available H\&E slides to reconstruct or predict HR gene-expression maps \cite{12}.  
Instead of grouping the literature by model archetype, we review it through the lens of three \textbf{key challenges}. For completeness, we cover all classic methods \cite{he2020integrating,bergenstraahle2022super,zhang2024inferring,tesla2023,controlnet2023} and explicitly point out how our work differs at the end of each subsection.

\noindent\textbf{Resolution Mismatch: From Spots to Subcellular Scale }Early studies confirmed a strong link between tissue morphology and gene expression.  
He~\textit{et~al.}\, \cite{he2020integrating} employed an ImageNet-pretrained DenseNet to regress the spot-level expression of 250 genes in breast cancer, demonstrating multi-gene prediction but inheriting the coarse spot grid.  
XFuse \cite{bergenstraahle2022super} mixed multi-scale latent variables of H\&E and ST through a down-sampling reconstruction loss, while iStar \cite{zhang2024inferring} introduced spatial priors into a Vision Transformer under weak supervision. Both still rely on LR ST labels and fail to capture pixel-level details. 

\noindent\textbf{Heterogeneous-Modality Alignment}
H\&E image translation must align two heterogeneous modalities: morphology and molecules.  
TESLA \cite{tesla2023} embeds both modalities into a unified graph and spreads information via graph convolutions; ControlNet \cite{controlnet2023} and Uni-ControlNet \cite{zhao2024uni} insert explicit conditioning branches into large diffusion models. Despite their success, these approaches usually fuse modalities by channel concatenation or simple addition and lack dynamic filtering of shared versus unique features, leading to blurred reconstructions in structurally complex tissues.

\noindent\textbf{Multi-Gene Synergy}
Gene expression exhibits strong synergy and complementarity; modeling each gene independently discards latent co-regulation.  
BayesSpace \cite{zb2022bayesspace} uses Bayesian statistics and spot adjacency to refine sub-spot inference, but ignores gene-level interactions.  
In MRI synthesis, DisC-Diff \cite{mao2023discdiff} deploys SE attention to weight each contrast channel globally, yet overlooks local differences. Video and multispectral methods such as MCCNet \cite{deng2021arbitrary} and GCRVFL \cite{altena2022improved} confirm the value of channel correlation but operate on global statistics only.  

% \subsection{Summary and Outlook}

% Current methods still suffer from \textbf{(i) resolution mismatch}, \textbf{(ii) heterogeneous-modality alignment}, and \textbf{(iii) multi-gene synergy}.  
% Our \textbf{HaDM-ST} tackles these gaps through a histology-assisted differential modeling methods, uniting the proposed HSD,CMSA and GDAL to maintain global morphology–expression consistency while restoring local details—paving the way toward clinically useful HR ST enhancement.
\vspace{-3mm}
\section{Methodology}
\label{sec:method}

\subsection{Problem Formulation and Overview}
\begin{figure*}[h]
\centering
\includegraphics[width=0.9\linewidth]{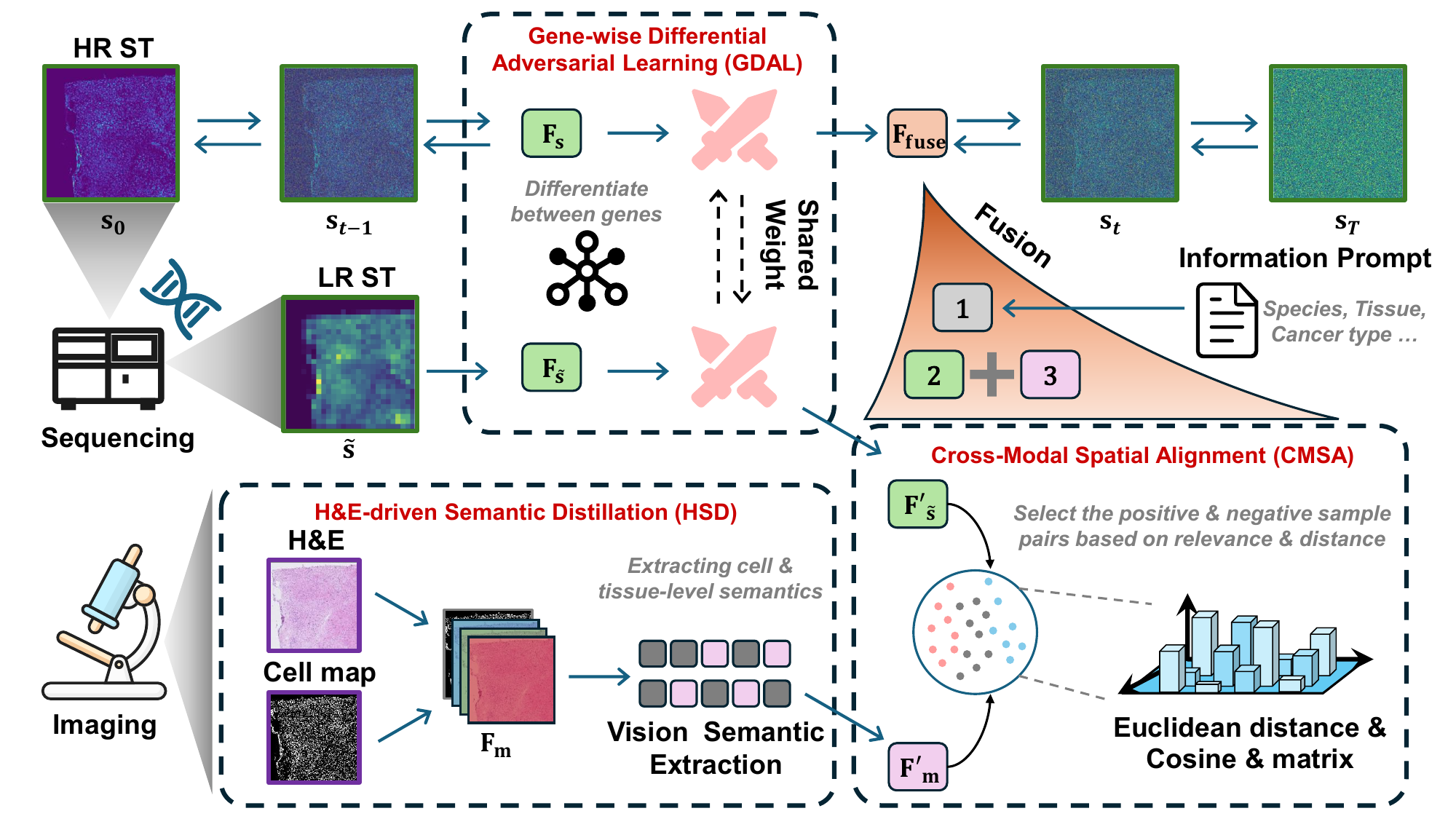}
\caption{Overall architecture of our histology-assisted differential modeling methods, comprising (A) the gene-wise differential adversarial learning module (GDAL), (B) the H\&E-driven semantic distillation module (HSD), and (C) the cross-modal spatial alignment module (CMSA), and we additionally include an information prompt to guide the reverse diffusion process for HR ST generation.}
\label{fram}
\vspace{-1.5em}
\end{figure*}

As shown in \textbf{Figure \ref{fram}},We propose a image translation method for ST Generation in histology-assisted differential modeling(HaDM-ST), which conditions on H\&E-stained histology images and LR ST measurements to reconstruct HR ST maps via the reverse diffusion process. Specifically, let $\tilde{\mathbf{s}}\!\in\!\mathbb{R}^{C\times H_{l}\times W_{l}}$ be a
LR ST tensor with $C$ gene channels,
and let $\mathbf{m}\!\in\!\mathbb{R}^{3\times H_{m}\times W_{m}}$ denote the
co-registered H\&E image of the same tissue section
($H_{m}\!\approx\!10H_{l},\,W_{m}\!\approx\!10W_{l}$ in practice).
Our goal is to synthesise a HR ST map
$\hat{\mathbf{s}}\!\in\!\mathbb{R}^{C\times H\times W}$,
where $H,W\!\gg\!H_{l},W_{l}$, by leveraging both
histology morphology and LR-ST measurements.

%Figure~\ref{fig:fram}summarises the framework.
\vspace{-3mm}
\subsection{Forward Stochastic Degradation}
\label{ssec:forward}
Following DDPM~ \cite{ho2020ddpm}, we denote the clean HR sample by
$\mathbf{s}_{0}$ and corrupt it over $T$ timesteps with a variance schedule
$\{\beta_{t}\}_{t=1}^{T}$:
\begin{equation}
q(\mathbf{s}_{1:T}\!\mid\!\mathbf{s}_{0})
=\prod_{t=1}^{T}
q(\mathbf{s}_{t}\!\mid\!\mathbf{s}_{t-1}),\qquad
q(\mathbf{s}_{t}\!\mid\!\mathbf{s}_{t-1})
=\mathcal{N}\!\bigl(
\mathbf{s}_{t};\sqrt{1-\beta_{t}}\,\mathbf{s}_{t-1},
\beta_{t}\mathbf{I}\bigr).
\label{eq:forward}
\end{equation}
Conveniently, $\mathbf{s}_{t}$ can be sampled in closed form as
$\mathbf{s}_{t}=\sqrt{\bar\alpha_{t}}\mathbf{s}_{0}
+\sqrt{1-\bar\alpha_{t}}\boldsymbol{\epsilon}$,
where $\bar\alpha_{t}\!=\!\prod_{i=1}^{t}(1-\beta_{i})$
and $\boldsymbol{\epsilon}\!\sim\!\mathcal{N}(\mathbf{0},\mathbf{I})$.
\vspace{-3mm}
%-----------------------------------------------------------------------
\subsection{Conditional Reverse Denoising}
\label{ssec:reverse}
At each timestep $t$, a step-adaptive condition vector

$\mathbf{c}_{t}=g_{t}\!\bigl(\psi(\mathbf{m}),\phi(\tilde{\mathbf{s}})\bigr)$
is formed by fusing morphology features
$\psi(\mathbf{m})$ (Sec.~\ref{ssec:hsd}) and aligned LR-ST features
$\phi(\tilde{\mathbf{s}})$ (Sec.~\ref{ssec:cmsa}).
The reverse transition is modelled as
\begin{equation}
p_{\theta}\!\bigl(\mathbf{s}_{t-1}\!\mid\!\mathbf{s}_{t},\mathbf{c}_{t}\bigr)
=\mathcal{N}\!\Bigl(
\mathbf{s}_{t-1};
\mu_{\theta}(\mathbf{s}_{t},\mathbf{c}_{t},t),\,
\sigma^{2}_{t}\mathbf{I}\Bigr),
\label{eq:reverse}
\end{equation}
where the mean is parameterised via
$\mu_{\theta}(\mathbf{s}_{t},\mathbf{c}_{t},t)=
\bigl(\mathbf{s}_{t}
-\tfrac{1-\alpha_{t}}{\sqrt{1-\bar\alpha_{t}}}\,
\epsilon_{\theta}(\mathbf{s}_{t},\mathbf{c}_{t},t)\bigr)\!
/\sqrt{\alpha_{t}}$,
and $\epsilon_{\theta}$ is a U-Net predicting the added noise.  
\textbf{Inference-time flexibility}: if $\tilde{\mathbf{s}}$ is unavailable,
$\phi(\tilde{\mathbf{s}})$ is omitted and $\mathbf{c}_{t}$ degrades gracefully
to $\psi(\mathbf{m})$.

\vspace{-3mm}
\subsection{H\&E-Driven Semantic Distillation (HSD)}
\label{ssec:hsd}

Due to the semantic discrepancy between H\&E images (tissue morphology) and ST data (gene expression), we design a multimodal fusion framework to bridge this gap. Let the H\&E image be denoted by \(\mathbf{I_{m}}\) and its corresponding cell‐segmentation map by \(\mathbf{I_{\mathrm{seg}}}\). We concatenate these two inputs and feed them into a Transformer network \(\mathcal{T}\) to obtain a high‐level semantic feature vector:
\begin{equation}
\mathbf{F_{m}} \;=\; \mathcal{T}\!\bigl(\mathrm{Concat}(\mathbf{I_{m}},\,\mathbf{I_{\mathrm{seg}}})\bigr).
\end{equation}

Furthermore, a cancer‐type prompt text is passed through a pretrained BERT model \(\mathcal{B}\) to yield an embedding vector \(\mathbf{E_{\mathrm{text}}}\), thereby incorporating biological priors that enhance the biological validity of the features:
\begin{equation}
\mathbf{E_{\mathrm{text}}} \;=\; \mathcal{B}\!\bigl(\mathrm{Prompt}_{\mathrm{cancer}}\bigr).
\end{equation}

The fusion of \(\mathbf{F_{m}}\) and \(\mathbf{E_{\mathrm{text}}}\) effectively reduces redundant visual information and more precisely guides the reconstruction of the high‐resolution ST map.
\vspace{-3mm}
\subsection{Cross-Modal Spatial Alignment (CMSA)}
\label{ssec:cmsa}
To address the spatial resolution and sampling‐position discrepancies between H\&E images and LR ST data, we design a feature alignment module based on contrastive learning. Let the H\&E features extracted by a UNet branch be denoted by \(\mathbf{F_{m}}\) and the LR ST features by \(\mathbf{F_{\tilde{s}}}\). We construct a cosine similarity matrix \(\mathbf{C}\) and a Euclidean distance matrix \(\mathbf{D}\):

\begin{equation}
C_{ij}
= \frac{\mathbf{F_{m,i}} \cdot \mathbf{F_{\tilde{s},j}}}
       {\|\mathbf{F_{m,i}}\|\,\|\mathbf{F_{\tilde{s},j}}\|},
\quad
D_{ij}
= \bigl\|\,\mathbf{F_{m,i}} - \mathbf{F_{\tilde{s},j}}\bigr\|.
\end{equation}

We then select sample pairs according to \(\mathbf{C}\): the top 30\% of region‐pairs by similarity are treated as positive samples, and the bottom 30\% as negative samples. On this basis, we integrate a cosine loss \(\mathcal{L}_{\mathrm{cosine}}\), an Euclidean loss \(\mathcal{L}_{\mathrm{euclidean}}\), and an InfoNCE mutual‐information loss \(\mathcal{L}_{\mathrm{InfoNCE}}\), weighting each term by coefficients \(\lambda_{1}\) and \(\lambda_{2}\), to form the overall contrastive loss:

\begin{equation}
\mathcal{L}_{\mathrm{contrast}}
= \mathcal{L}_{\mathrm{cosine}}
+ \lambda_{1}\,\mathcal{L}_{\mathrm{euclidean}}
+ \lambda_{2}\,\mathcal{L}_{\mathrm{InfoNCE}}.
\end{equation}

By minimizing \(\mathcal{L}_{\mathrm{contrast}}\), we ensure precise spatial and semantic alignment of cross‐modal features.
\vspace{-3mm}
\subsection{Gene-wise Differential Adversarial Learning (GDAL)}
\label{ssec:gdal}
Considering the complex co‐regulatory relationships inherent in true gene expression profiles, we designed a fine-grained channel-specific difference modeling module based on a graph neural network to precisely capture inter‐channel discrepancies. Specifically, we represent each gene channel as a node in a co‐expression graph \(G=(V,E)\), where the edge weight between nodes is computed from gene‐expression correlations . Denoting the feature vector of node \(v\) at layer \(l\) by \(H_v^{(l)}\), we perform feature propagation through a GNN to obtain context‐aware node embeddings:
\begin{equation}
H_v^{(l+1)} 
= \sigma\!\Bigl(\sum_{u\in\mathcal{N}(v)} a_{vu}^{(l)} \,W^{(l)}\,H_u^{(l)}\Bigr),
\end{equation}
where \(\sigma\) is the activation function, \(W^{(l)}\) is the learnable weight matrix at layer \(l\), \(a_{vu}^{(l)}\) denotes the dynamic edge weight from node \(u\) to \(v\), and \(\mathcal{N}(v)\) is the neighborhood of \(v\). Finally, these node features are fused with the H\&E and low‐resolution ST features, enabling channel-level gene-wise differentiation and thus further enhancing the realism and biological interpretability of the reconstructed high‐resolution ST data.
\begin{figure}[!t]
  \centering

    \centering
    \includegraphics[width=0.8\linewidth]{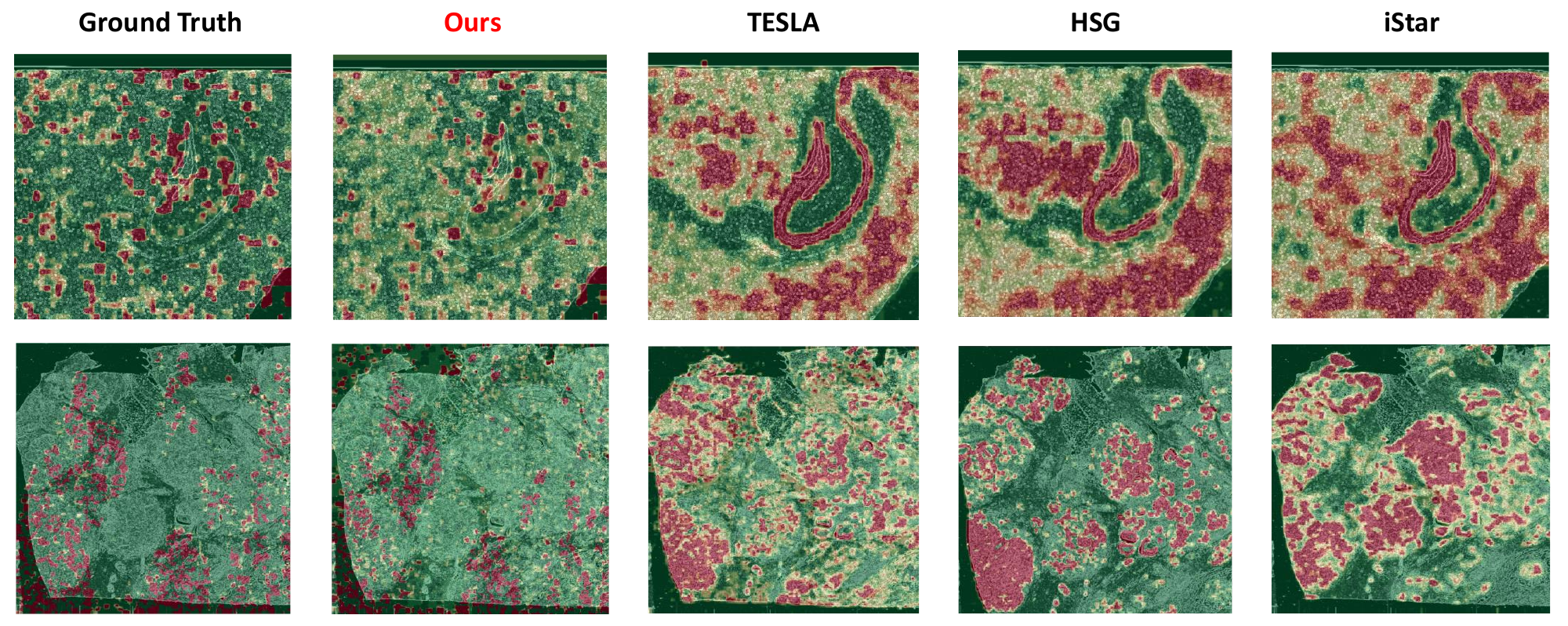}
  \caption{Local structural Similarity index measure (SSIM)-based spatial alignment evaluation between ST and H\&E histology. Gradient-enhanced H\&E images are overlaid with semi-transparent RdYlGn heatmaps of sliding-window SSIM , where red denotes low alignment, yellow moderate alignment, and green high alignment. The upper panel corresponds to the mouse brain, and the lower panel to the human breast.}
  \label{reli}
\end{figure}

\begin{figure}[!t]
  \centering

    \centering
    \includegraphics[width=0.9\linewidth]{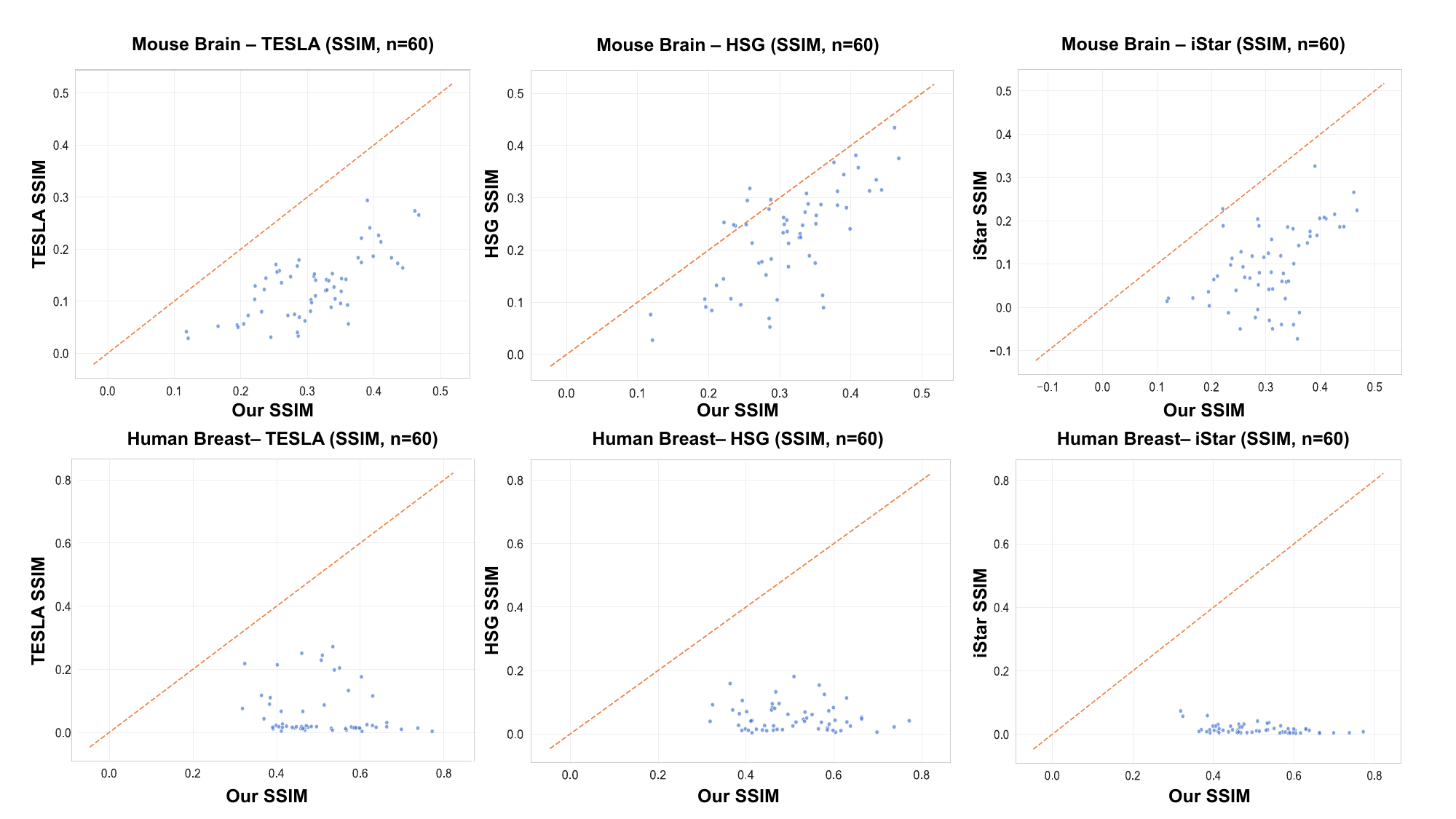}
  \caption{Comparison of SSIM performance across multiple algorithms on the mousebrain and human breast Xenium datasets.}
  \label{sandian}
\end{figure}

\section{Experiments \& Results}

\subsection{Datasets and Gene Selection}
We benchmark \textbf{HaDM-ST} on two publicly available Xenium spatial-transcriptomics cohorts: 
Mouse Brain and Human Breast \cite{1}. 
For each cohort, we curate 200 highly variable genes; removing overlaps yields 
120 unique genes. 
In total, we process 514 paired H\&E slides and 
61\,680 ST image tiles. 
For the Human Breast cohort, 85 slides (17\,000 tiles) are randomly divided, 
with 80\% used for training and 20\% for testing. 
Each H\&E and HR ST tile is resized to \textbf{\(256\times256\)} pixels 
(10\,\textmu m per pixel), whereas LR ST maps are down-sampled to 
\(26\times26\) pixels (100\,\textmu m per pixel).

\vspace{-1em}
\subsection{Implementation Details}
All experiments are conducted on two NVIDIA RTX V100 GPUs (32\,GB memory). 
The network is trained for 20\,000 epochs with a batch size of 4, 
an initial learning rate of \(1\times10^{-4}\), and the AdamW optimiser \cite{loshchilov2017decoupled} 
with weight decay. 
Following the sampling policy of \cite{dhariwal2021diffusion}, we use 
1\,000 diffusion timesteps for both the forward and reverse processes. 
Key hyper-parameters are listed in Supplementary Table~I, and all settings 
are tuned on the validation set.

%%%%%%%%%%%%%%%%%%%%%%%%%%%%%%%%%%%%%%%%%%%%%%%%%%%%%%%%%%%%%%%
\vspace{-1em}
\subsection{Performance evaluation}

\noindent\textbf{Quantitative comparison:} 
We compare our model with three SOTA methods, including TESLA \cite{tesla2023}, HiStoGene(HSG) \cite{2021histogenepang} and iStar \cite{zhang2024inferring}(conference version of our method). Among these, TESLA, HSG and istar are specially designed for ST SR.To ensure a fair comparison, all methods utilize both H\&E images and LR ST maps to enhance ST maps. 

%2021histogenepang
%Xfuse \cite{bergenstraahle2022super}

We use two metrics for model evaluation: structure similarity index measure (SSIM), root MSE (RMSE)). As shown in \textbf{Table \ref{tab:performance-10x}}, our method achieves the best performance. It improves SSIM by at least 0.0370 and reduces RMSE by 0.053 on the mouse brain-Xenium dataset, and improves SSIM by at least 0.4008 and reduces RMSE by 0.0528 on the human breast-Xenium dataset, demonstrating its effectiveness in integrating H\&E features and gene expressions for ST SR.

As we can see in \textbf{Fig.~\ref{reli}}, our local SSIM–based alignment maps exhibit predominantly green regions across both the mouse brain–Xenium and human breast–Xenium datasets, indicating high spatial concordance between ST measurements and H\&E histology. These results demonstrate that our SSIM-driven framework reliably captures fine‐scale morphological correspondences, thereby providing a solid quantitative foundation for downstream ST analyses.

Further, compared to all SOTA methods specially designed ST SR, our approach excels in reconstructing structural information, As we can see \ref{sandian} presents SSIM scatter comparisons between our method (x-axis) and three state-of-the-art baselines—TESLA, HSG, and iStar—on both the mouse brain Xenium (top row) and human breast Xenium (bottom row) datasets. Each panel plots SSIM over 60 gene samples, with the dashed $y = x$ line indicating equal performance. In all six plots, the majority of points lie below the diagonal, demonstrating that our approach consistently attains higher structural similarity and thus superior fidelity across both tissue types. These significant gains could be due to our designs for extracting spatial patterns from both H\&E images and ST maps. 
%In further ablation studies (Section ???????????????????????????\ref{ablation}),
Notably, the ST SR task remains highly challenging due to the remarkable heterogeneity in spatial gene expression \cite{miller2021characterizing}, leading to complex data distributions and severe class imbalance.

\begin{table}[!t]
  \centering
  \scriptsize
  \caption{Performance comparisons on two datasets with 10$\times$ enlargement scales. Bold numbers indicate the best results.}
  \label{tab:performance-10x}

  %=== 子表 (a) Xenium mouse ===%
  \begin{minipage}[t]{0.48\textwidth}
    \centering
    \begin{tabular}{l|cc}
      \toprule
      Approach & RMSE   & SSIM    \\
      \midrule
      TESLA     & 0.2489 & 0.1373  \\
      iStar     & 0.3088 & 0.0995  \\
      HSG     & 0.2000 & 0.2648  \\
      \midrule
      \textbf{Ours} & \textbf{0.1630} & \textbf{0.3184} \\
      \bottomrule
    \end{tabular}
    \\[0.5ex]
    {\footnotesize (a) Mouse brain}
  \end{minipage}%
  \hfill
  %=== 子表 (b) Xenium human breast ===%
  \begin{minipage}[t]{0.48\textwidth}
    \centering
    \begin{tabular}{l|cc}
      \toprule
      Approach & RMSE   & SSIM    \\
      \midrule
      TESLA     & 0.3302 & 0.0655  \\
      iStar     & 0.3071 & 0.0486  \\
      HSG       & 0.2832 & 0.0533  \\
      \midrule
      \textbf{Ours} & \textbf{0.2304} & \textbf{0.4663} \\
      \bottomrule
    \end{tabular}
    \\[0.5ex]
    {\footnotesize (b) Human breast}
  \end{minipage}

\end{table}

\begin{figure}[!t]
  \centering
  % 左图：宽度占 49.5% 全文宽度
  \begin{subfigure}[t]{0.495\textwidth}
    \centering
    \includegraphics[width=\linewidth]{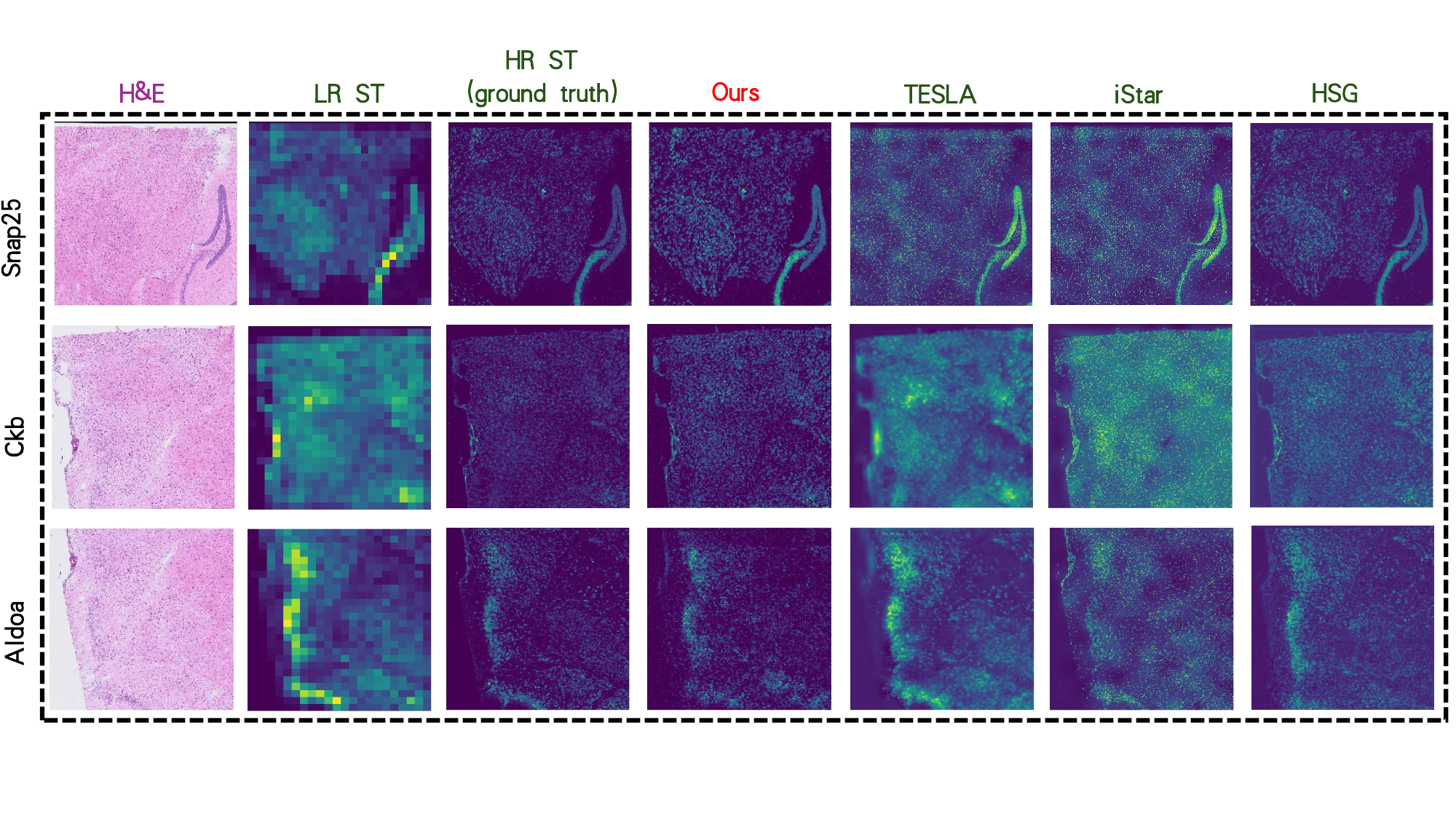}
  \end{subfigure}%
  % 只留 0.5% 宽度的间隔
  \hspace{0.005\textwidth}%
  % 右图：宽度占 49.5% 全文宽度
  \begin{subfigure}[t]{0.495\textwidth}
    \centering
    \includegraphics[width=\linewidth]{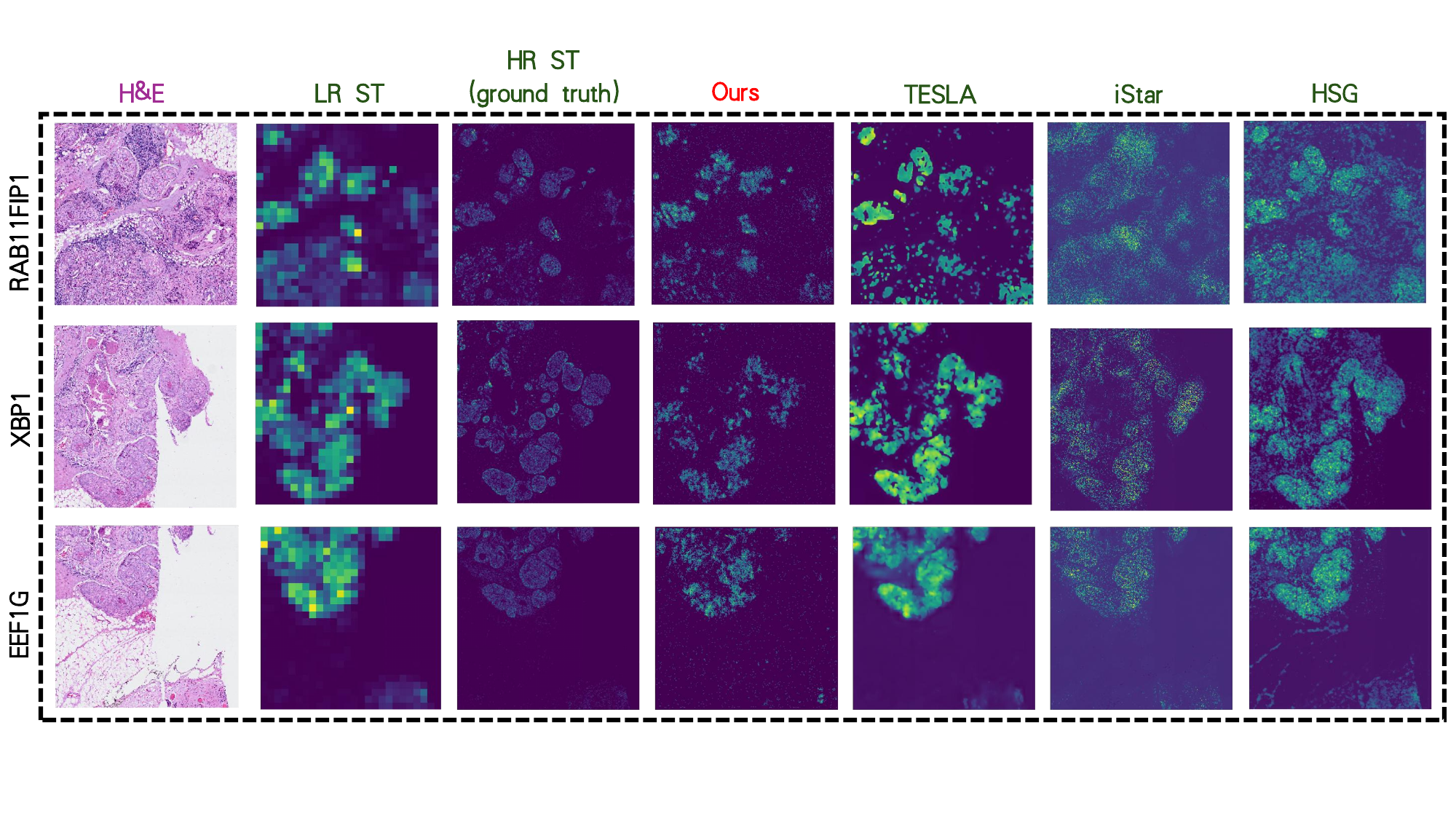}
  \end{subfigure}

  \caption{Visual comparisons on the mouse brain dataset (Aldoa, Ckb and Snap25) and on the human breast dataset (RAB11FIP1, XBP1 and EEF1G).}
  \label{fig:visual-comparison-10x}
\end{figure}

\noindent\textbf{Visual comparison.}
\textbf{Fig. \ref{fig:visual-comparison-10x}} presents the restoration results of our method alongside the three best-performing ST SR methods on both datasets. Our approach consistently outperforms others, generating HR ST maps with sharper edges and finer details. 
% Additional visual comparisons can be found in Supplementary Fig. 3.

\section{Conclusion}

We propose a novel diffusion-based framework that integrates semantic distillation, cross-modal spatial alignment, and gene-wise adversarial learning to improve the accuracy and interpretability of histology-to-transcriptomics image translation. Quantitative and qualitative experiments demonstrate the effectiveness of our approach in three key aspects: extracting expression-relevant semantics from H\&E images, achieving precise spatial co-registration between modalities, and modeling fine-grained gene expression patterns across channels. Our method provides a robust foundation for advancing ST applications in precision medicine and offers new insights into the molecular mechanisms underlying tissue organization and disease progression.

\bibliographystyle{ieeetr}   % 样式 plain、ieeetr、unsrt、apalike...
\bibliography{ref}
% \bibliography{refs}
\end{document}